\documentclass[epj]{svjour}
\usepackage{graphicx}
\usepackage{amsmath}
\def\eq#1{Eq.~(\ref{#1})}
\newcommand{\eqs}[2]{Eqs.~(\ref{#1}) and (\ref{#2})}
\def\fig#1{Fig.~\ref{fig:#1}}

\def\sec#1{Sec.~\ref{#1}}

\newcommand{\erf}{\operatorname{erf}}
\newcommand{\Erf}[1]{\erf\left(#1\right)}
\def\12{\frac{1}{2}}
\def\kt{k_BT}
\newcommand{\avg}[1]{\langle #1\rangle}
\newcommand{\Exp}[1]{\exp\left(#1\right)}
\newcommand{\e}[1]{e^{#1}}

\newcommand{\FIG}[3]{
\begin{figure*}[t]
\includegraphics[width=\textwidth]{#2}
\caption{\label{fig:#1}#3}
\end{figure*}}

\begin{document}
\title{Impact of receptor-ligand distance on adhesion cluster stability}
\author{Thorsten Erdmann 
\thanks{\emph{Present address:} FOM Institute for Atomic and Molecular Physics, Kruislaan 407, 1098 SJ Amsterdam, The Netherlands}
\and Ulrich S. Schwarz%
}                     % Do not remove
\institute{University of Heidelberg, Im Neuenheimer Feld 293, D-69120 Heidelberg, Germany}
\date{Received: date / Revised version: date}
% The correct dates will be entered by Springer
%
\abstract{
  Cells in multicellular organisms adhere to the extracellular matrix
  through two-dimensional clusters spanning a size range from very few
  to thousands of adhesion bonds. For many common receptor-ligand
  systems, the ligands are tethered to a surface via polymeric spacers
  with finite binding range, thus adhesion cluster stability crucially
  depends on receptor-ligand distance. We introduce a one-step master
  equation which incorporates the effect of cooperative binding
  through a finite number of polymeric ligand tethers. We also derive
  Fokker-Planck and mean field equations as continuum limits of the
  master equation. Polymers are modeled either as harmonic springs or
  as worm-like chains. In both cases, we find bistability between
  bound and unbound states for intermediate values of receptor-ligand
  distance and calculate the corresponding switching times. For small
  cluster sizes, stochastic effects destabilize the clusters at large
  separation, as shown by a detailed analysis of the stochastic
  potential resulting from the Fokker-Planck equation.
 % end of PACS codes
} %end of abstract
\maketitle

\section{Introduction}

Cells adhere to the extracellular matrix or to each other through a
multitude of weak interactions. The transient character of their
adhesions allows cells to adapt quickly to changes in their
environment. In particular, cell adhesion has to be transiently
down-regulated during important physiological processes like migration
or division. For tissue cells, the main receptors for cell-matrix
adhesion are integrins, which are linked on the extracellular side to
ligands like fibronectin and on the cytoplasmic side to the actin
cytoskeleton \cite{c:bers99}. This provides structural integrity
between extracellular matrix and cytoskeleton, which is important
because cell-matrix adhesion clusters usually have to operate under
considerable mechanical load. The behaviour of single receptor-ligand
bonds under force was first discussed by Bell \cite{c:bell78}, who
suggested that bond lifetime is reduced exponentially by an applied
load. This concept has been impressively confirmed and extended by
dynamic force spectroscopy
\cite{c:evan97,c:merk99,c:evan01a,c:merk01,c:raib06}, which showed
that binding strength can only be defined in a dynamical context.  In
particular, it has been shown both theoretically and experimentally
that for linearly increasing load, binding strength increases
linearly with the logarithm of loading rate. If loading occurs through
a soft polymeric linker, binding strength is decreased compared to
loading at the same speed but through a rigid linker \cite{c:evan99}.
In practise, cell adhesion does not work with single bonds, but with
clusters of bonds. It has been noted earlier that the distribution of
load over the receptors in a finite-sized cluster induces non-trivial
cooperativity between adhesion bonds: as one bond is disrupted, the
load on the remaining bonds increases and cluster stability diminishes
\cite{c:bell78,c:seif00,a:Seifert2002,uss:erdm04a}.  Likewise, as a
new bond is established, the other bonds feel less force and cluster
stability is enhanced.

For many common receptor-ligand systems, the ligands are tethered to a
surface through polymeric spacers. This reduces the disturbance of
ligand structure to preserve its specificity and allows exploration of
space for receptors so that the effective affinity of surfaces covered
with specific bonds is increased. For example, it has been shown
recently in a macroscopic shearing experiment for streptavidin-biotin
bonded beads that bonding is enhanced if ligands are tethered with
polymeric spacers \cite{c:lebo06}. Thus the distance between ligands
and receptors is an important determinant for specific adhesion. When
cells adhere to and eventually spread on a surface, the distance
between the ligand- and receptor-carrying surfaces decreases from the
$\mu\rm m$- to the $\rm nm$-range on the time scale of minutes
\cite{a:CohenEtAl2004}. The final receptor-ligand distance for
integrins is of the order of $15-20\ \rm nm$, which is bridged by the
polymer spacer carrying the ligand.

During recent years, the distance-dependent binding of tethered
ligands has been investigated experimentally as well as theoretically.
In experiments using the surface force apparatus and high-affinity
streptavidin-biotin bonds \cite{a:WongEtAl1997,a:JeppesenEtAl2001}, it
has been shown that binding depends on rare, strongly extended
conformations of the spacers and that the onset of binding is followed
by a very fast increase of the fraction of bound tethers. Binding of
tethered ligands to receptors has been described using a combination
of Monte Carlo simulations and reaction-diffusion theory.  Simulations
of pearl-bead chains confined between two walls were used to determine
the force-extension relation for the polymers and to derive a
potential landscape for the movement of the ligands.  Later reversibel
tethered bonds have been treated as deep but finite potential wells in
the polymer potential landscape derived from simulations and used in
the reaction-diffusion equations \cite{a:MoreiraEtAl2003,a:MoreiraMarques2004}.
Moreover, the effect of changing receptor-ligand distance with a
prescribed velocity has been discussed. It was concluded that kinetic
effects are most important for strong binders with large affinity
while thermodynamic equilibrium dominates for weak, reversible binders
with small affinity. Recently, a theoretical treatment has also been
given for receptor-ligand binding between curved surfaces, where
different bonds are not equivalent for geometrical reasons \cite{c:moor06}.

It is important to note that these theoretical studies have been
conducted in the framework of a mean-field description for a large
numbers of independent, non-cooperative bonds. No theoretical
treatment has been presented yet for the impact of receptor-ligand
distance on the stability and dynamics of finite-sized clusters with
cooperative bonds.  Experimentally, it is well known that the lateral
arrangement of the integrins and therefore the integrin cluster size
is regulated by cytoplasmic proteins like talin and $\alpha$-actinin,
which can bind both to the integrins and to the actin cytoskeleton
\cite{c:geig01a}.  Using image correlation microscopy, it has been
shown for migrating cells that the integrins which are not yet
organized in adhesions are already preclustered with an average
cluster size of three to four \cite{c:wise04}. With a measured area
density of few hundreds of integrins per $\mu m^2$, this corresponds
to a lateral distance well above 100 nm between the different
mini-clusters. As the adhesion contacts nucleate and grow, integrins
are increasingly clustered, until they approach a density of 1.000 -
10.000 per $\mu m^2$, corresponding to a lateral distance of 10 - 30
nm. The progression from very few to thousands of integrins per
cluster suggests that finite size effects might be highly relevant in
the stabilization of initial cell adhesion.

In this paper, we theoretically study the effect of feedback and
cooperativity on the distance-dependent receceptor-ligand binding
dynamics in finite-sized adhesion clusters. To this end, we use a
one-step master equation, which has been used before to study the
adhesion cluster stability under mechanical load, both for constant
force \cite{uss:erdm04a,uss:erdm04c} and linearly rising force
\cite{uss:erdm04b}. Here we extend this framework to include the
effect of receptor-ligand distance. We start in \sec{sec:model} by
introducing the appropriate one-step master equation. In addition, we
derive two corresponding continuum descriptions, namely a
Fokker-Planck equation and a deterministic differential (mean field)
equation. Our modeling framework can be used for any spatial
distribution of the ligand in the direction normal to the
substrate to which it is tethered. For simplicity, we start with a
harmonic tether potential (\textit{spring model}), which is the first
order approximation for all polymer models at small extensions. In
\sec{sec:bifurcation}, we analyze the stationary states of the mean
field equation and derive one-parameter bifurcation diagrams which
show that the receptor-ligand dynamics of adhesion clusters leads to
bistability between bound and unbound states. In
Secs.~\ref{sec:meqstat} and \ref{sec:meqdyn} we discuss stationary and
dynamic properties of the master equation, respectively. In
particular, we find that large adhesion clusters are stabilized in the
bound state due to very large switching times to the unbound state. In
\sec{sec:wlc} we combine our conceptual framework with the worm-like
chain model for semiflexible polymers to study the effect of finite
polymer contour length.  We conclude in \sec{sec:conclusion} by
discussing some biological applications of our results. A short report
on our main results regarding the spring model has been given before
\cite{uss:erdm06a}.

\section{Model}\label{sec:model}

\subsection{Master equation and continuum limits}

\FIG{cartoon}{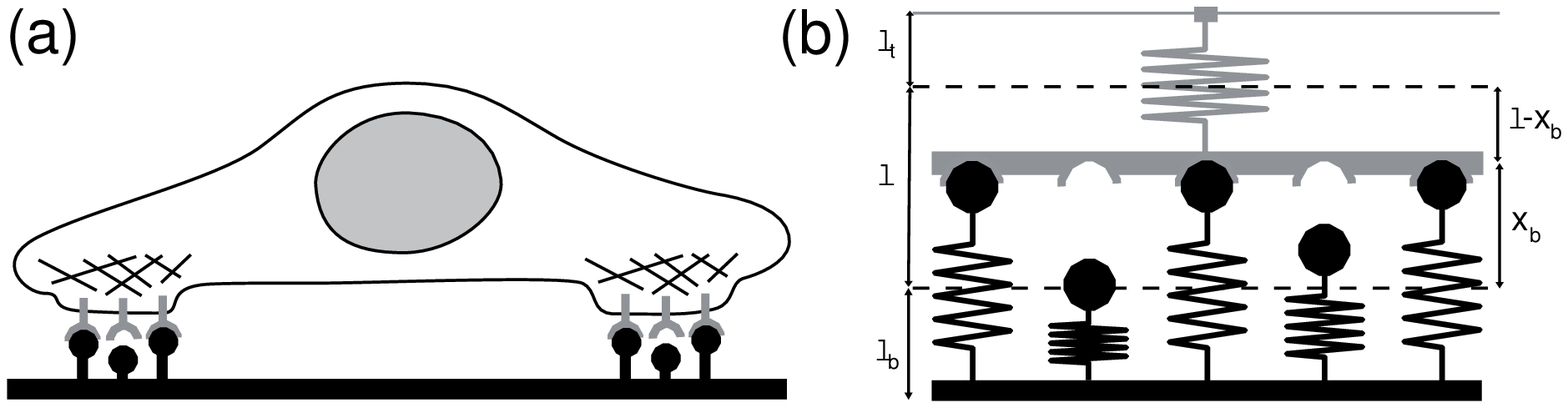}{(a) Cartoon of a cell adhering to a substrate
  through two sites of adhesion. Each adhesion corresponds to an
  elastic deformation in regard to the average cell-substrate
  distance.  (b) Simple mechanical model for a site of adhesion. The
  cluster consists of $N_t$ bonds (here $N_t = 5$).  At a given time
  $t$, $i$ of these bonds are closed (here 3) while $N_t - i$ (here 2)
  are open. The force transducer at the top (cell envelope) and the
  ligand tethers at the bottom are modeled as harmonic springs, with
  rest lengths $\ell_t$ and $\ell_b$ and spring constants $k_t$ and
  $k_b$, respectively. The equilibrium positions of the springs
  (dashed lines) are separated by the receptor-ligand distance $\ell$.
  The extensions of the bond springs and the transducer are denoted by
  $x_b$ and $x_t = \ell - x_b$, respectively.}

\fig{cartoon} shows a schematic representation of our model for an
adhesion cluster. The mechanical properties of the force transducer
holding the receptors and of the polymeric tethers holding the ligands
are represented by harmonic springs. We consider a situation in which
$N_t$ receptor-ligand pairs are arranged in parallel between the planar
force transducer and the substrate. At a given time, each of these
adhesion bonds can either be open or closed. All bonds are considered to
be equivalent so that the state of the adhesion cluster is characterized
by the number of closed bonds $i$ alone. The number of open bonds is
$N_t - i$. Because $i$ ranges from $i=0$ (completely unbound state) to
$i=N_t$ (completely bound state), our model has $N_t+1$ discrete states.
Every bond changes its state (rebinds or ruptures) through thermally
activated, stochastic transitions. The stochastic rates for these
transitions will be specified below.  Therefore, the stochastic variable
$i$ changes by discrete steps $\pm 1$ (\textit{one-step process}). The
time dependence of the functions $\left\{p_i(t)\right\}_{i=0}^{N_t}$
representing the probability that $i$ bonds are closed at time $t$ are
described by a one-step master equation which has the general form
\begin{equation}\label{MasterEquation}
\frac{dp_i}{dt} = r_{i+1} p_{i+1} + g_{i-1} p_{i-1} - \left\{r_i + g_i \right\} p_i\,.
\end{equation}
The forward rates $g_i$ for the formation and the reverse rates $r_i$
for the rupture of a closed bond have to be specified for all states $0
\le i \le N_t$.  They have the general form $r_i = i k_{off}(i)$ and
$g_i = (N_t - i) k_{on}(i)$.  The reverse rate is the product of the
number of closed bonds and the single bond off-rate $k_{off}(i)$ because
the closed bonds rupture independently and at a given time any of them
could be the next to break. The forward rate is the product of the
number of open bonds $(N_t-i)$ and the single bond on-rate $k_{on}(i)$
because the open bonds are independent and any one of them could be the
next to bind. Both, off- and on-rate will in general depend on $i$. The
form of the transition rates implies that the system is constrained to
the interval $0\le i \le N_t$ by two natural, reflecting boundaries at
$i=0$ and $i=N_t$, provided the single bond rates are finite at these
boundaries.

If the transition rates are continuous functions of $i$, that is $g_i =
g(i)$ and $r_i = r(i)$, it is useful to introduce a mean-field
description for the average number of closed bonds, $N(t) = \avg{i} =
\sum_{i = 1}^{N_t} i p_i(t)$. The time derivative of $N(t)$ follows from
the master equation \eq{MasterEquation} in terms of averages over the
transition rates. Dragging the average into the argument, it can be
approximated as
\begin{equation}\label{DeterministicEquation}
\frac{dN}{dt} = \avg{g(i)} - \avg{r(i)} \approx g(\avg{i}) - r(\avg{i}) = g(N) - r(N)\,.
\end{equation}
This procedure yields an ordinary differential equation for $N$ which in
the following we term the \textit{mean field equation}. It is an exact
equation for the mean number of bonds $N$ if the transition rates are
linear functions in $i$. Otherwise, it is only valid in the limit of
large system size $N_t$.

To assess the role of fluctuations in the continuum limit, we derive a
Fokker-Planck equation from the master equation. To achieve this, the
one-step master equation \eq{MasterEquation} is written in operator form
\cite{b:VanKampen2003}
\begin{equation}\label{OperatorMaster}
\frac{\partial p(i,t)}{\partial t} = \left\{(E-1)r(i) + (E^{-1}-1)g(i)\right\}p(i,t)
\end{equation}
where the operators $E$ and $E^{-1}$ act on the index $i$ as
\begin{equation}
Ef(i) := f(i+1) \quad\text{and}\quad E^{-1}f(i) := f(i-1)\,.
\end{equation}
In a continuum description for $i$ these can be expanded in a Taylor series,
\begin{equation}\label{OperatorExpansion}
E      = 1 + \sum_{n=1}^{\infty}\frac{  1   }{n!}\partial_i^n \quad\text{and}\quad
E^{-1} = 1 + \sum_{n=1}^{\infty}\frac{(-1)^i}{n!}\partial_i^n \,.
\end{equation}
Inserting \eq{OperatorExpansion} in \eq{OperatorMaster} leads to the Kramers-Moyal
expansion \cite{b:Risken1989}
\begin{equation}\label{KramersMoyal}
\frac{\partial p(i,t)}{\partial t} = \left\{-\sum_{n=0}^{\infty}\frac{\partial_i^{2n+1}}{(2n+1)!}A(i) 
                           +\sum_{n=1}^{\infty}\frac{\partial_i^{2n}}{2n!}      D(i)\right\}p(i,t)\,,
\end{equation}
in which the Kramers-Moyal coefficients are defined by
\begin{equation}
A(i) = g(i) - r(i)  \quad\text{and}\quad  D(i) = g(i) + r(i)\,.
\end{equation}
The Kramers-Moyal expansion \eq{KramersMoyal} is an exact continuum
representation of the master equation. Terminating the infinite series after second
order yields the Fokker-Planck equation
\begin{equation}\label{FokkerPlanckEquation}
\frac{\partial p(i,t)}{\partial t} = \left\{-\partial_i A(i)+\frac{1}{2}\partial_i^2 D(i)\right\}p(i,t)
\end{equation}
where $A(i)$ is the drift and $D(i)$ the diffusion coefficient, which determine the
short time behavior of the first and second centered moments, respectively:
\begin{align}
A(i(t_0)) &= \lim_{\Delta t \to 0}\frac{\langle  i(t_0 + \Delta t) - i(t_0) \rangle}{\Delta t}\\
\intertext{and}
D(i(t_0)) &= \lim_{\Delta t \to 0}\frac{\langle (i(t_0 + \Delta t) - i(t_0))^2 \rangle}{\Delta t}\,.
\end{align}
By deriving the differential equations for average and variance directly
from the master equation, one can verify that the left hand sides are
the derivatives $dN/dt$ and $d\sigma^2/dt$ for the initial conditions
$N(t_0) = i(t_0)$ and $\sigma^2(t_0) = 0$ at time $t = t_0$
\cite{b:VanKampen2003,uss:erdm04c}. The Fokker-Planck equation thus
describes the average and the variance correctly. However, the centered
moments of higher order are not correctly represented due to the
truncation procedure.

In general, a continuous description of a one-step process through the
mean field equation \eq{DeterministicEquation} or the Fokker-Planck
equation \eq{FokkerPlanckEquation} will be valid if the variation of
transition rates and probability distribution is small over the step
size $\Delta i = \pm 1$. Below we will use the mean field equation for
a bifurcation analysis in order to demonstrate that our system is
bistable. Because the Fokker-Planck equation results from a truncated
Kramers-Moyal expansion, it cannot be used to describe the full
dynamics. For a bistable system it has been shown before that the
diffusion approximation \eq{FokkerPlanckEquation} leads to an error in
the stationary probability distribution and an overestimation of the
transition rates between the coexisting states
\cite{a:HaenggiEtAl1984}.  However, this error is small for
sufficiently smooth transition rates. Moreoever, the extrema of the
stationary distribution are correctly described by the diffusion
approximation of the Fokker-Planck equation as used here. Therefore we
will use it below to investigate how the stationary states are
affected by thermal fluctuations which are not included in the mean
field approximation. In order to describe the full stochastic
dynamics, we will use the original master equation
\eq{MasterEquation}. Alternatively, an improved Fokker-Planck
description could be used as explained in
Ref.~\cite{a:HaenggiEtAl1984}.

\subsection{Transition rates}\label{TransitionRates}

\subsubsection{Reverse rate}

The bond dissociation dynamics is characterized by the reverse rate
$r(i)$ and strongly depends on the forces acting in the cluster. We
start with the simplest possible model as suggested by the cartoon in
\fig{cartoon}, that is the transducer and ligand tethers are modeled
as harmonic springs with rest lengths $\ell_t$ and $\ell_b$ and spring
constants $k_t$ and $k_b$, respectively. Because the receptors are
usually firmly attached to the actin cytoskeleton, deformation of the
transducer requires a local deformation of the whole cell membrane as
shown in the cartoon of \fig{cartoon}(a).  Therefore the stiffness
$k_t$ of the transducer is the combined stiffness of plasma membrane
and cell cortex. The extensions $x_b$ for a bound ligand and $x_t$
for the transducer satisfy the relation $x_b + x_t = \ell$, where
$\ell$ is the receptor-ligand distance in the completely dissociated
state.  In the following, we will treat the relaxed or unloaded
receptor-ligand distance $\ell$ as a parameter which has been fixed
externally, e.g.\ by the average cell-substrate distance or in an
experiment with the surface forces apparatus.  Mechanical equilibrium
requires
\begin{equation}\label{MechanicalEquilibrium}
i F_b = F_t = k_t x_t = k_t (\ell-x_b) = k_t \left(\ell- F_b / k_b\right)
\end{equation}
so that
\begin{equation}\label{Force_i}
F_b(i) = k_bx_b(i) = \frac{k_b \ell}{1 + i (k_b/k_t)}\,.
\end{equation}
The dissociation rate of a bond under force is described by the Bell
model as $k_{off} = k_0 \Exp{F_b / F_0}$, where $k_0$ is the unstressed
off-rate and $F_0$ is the bond's internal force scale \cite{c:bell78}.
The Bell model can be rationalized in the framework of Kramers' theory
for the escape over a sharp energy barrier \cite{c:evan97}. The reverse
rate $r(i)$ for the one-step master equation now follows as
\begin{equation}\label{ReverseRate}
r(i) =  i k_0 \Exp{F_b(i)/F_0}\,.
\end{equation}
When a bond ruptures, force is redistributed according to \eq{Force_i}
over the smaller number of bonds so that the load on the remaining bonds
increases. As a consequence, the extension of the remaining tethers
increases which increases the force even further. A decrease of the
number of closed bonds thus increases the off-rate for the single bonds
and makes rupture of a further bond more likely. Thus the reverse rate
from \eq{ReverseRate} describes a positive feedback mechanism for
rupture which results from cooperativity in load sharing.

\subsubsection{Forward rate}

The bond association dynamics is characterized by the forward rate
$g(i)$ and is strongly determined by the receptor-ligand distance
which has to be bridged for the formation of a new bond. In the
cartoon of \fig{cartoon}, the ligands are attached to springs and can
move between the ligand-coated surface at the bottom and the
transducer surface at the top. Hence, they move in a truncated
harmonic potential, which is
\begin{equation}
U(x) = \frac{k_b}{2}x^2 \quad\text{if\, $-\ell_b \le x \le x_b$}
\end{equation}
and $U(x) = \infty$ otherwise. For a finite temperature $T$, the
probability density for the ligand to be at position $x \in
[-\ell_b,x_b]$ reads
\begin{equation}\label{density}
\rho(x) =  \frac{1}{Z} \Exp{-U(x)/\kt} = \frac{1}{Z} \Exp{-k_bx^2/2 \kt}
\end{equation}
with the partition sum
\begin{equation}\label{partitionsum}
Z = \left[\frac{\pi\kt}{2k_b}\right]^{\12}
         \left\{\Erf{\left[\frac{k_b\ell_b}{2\kt}\right]^{\12}}
              + \Erf{\left[\frac{k_bx_b^2}{2\kt}\right]^{\12}}\right\}\, .
\end{equation}
Here $\Erf{x} = (2/\sqrt{\pi}) \int_0^{\infty} \Exp{-t^2} dt$ is the
error function. The binding process can conceptually be divided into
two steps.  First, ligand and receptor have to come sufficiently close
to form an encounter complex
\cite{r:eige74,c:berg77,r:shou82,c:schr02}. Second, this entangled
state has to react to form the final complex. For a stationary
distribution of tethers, the first step is limited by $\rho(x_b)$, the
probability of the ligand to be close to the transducer surface. The
second step is described by the on-rate $k_{on}$ for the case that
ligand and receptor are sufficiently close within the binding radius
$\ell_{bind}$. The forward rate $g(i)$ for the one-step master
equation \eq{MasterEquation} thus reads
\begin{equation}\label{ForwardRate}
g(i) = k_{on}(N_t - i)\rho(x_b)\,.
\end{equation}
When an open bond rebinds, force is redistributed over the larger
number of bonds according to \eq{Force_i} and the load on the single
bonds decreases. This reduces the extension of the bound tethers and
increases the density of free ligands close to the transducer. An
increase of the number of closed bonds thus increases the on-rate for
the single bonds and makes rebinding of a further bond more likely.
Thus the forward rate from \eq{ForwardRate} describes a positive
feedback mechanism for binding which results from cooperativity in
the formation of encounter complexes.

\subsection{Adiabatic assumption}

In the derivation of reverse and forward rates it was assumed that
relaxation of transducer and polymer tethers to mechanical equilibrium
after a change in $i$ is fast compared to rupture and rebinding of
adhesion bonds. This assumption is a prerequisite for the definition
of discrete states described by the number of closed bonds alone and
hence for the validity of the master equation. Experimentally it has
been found for a biomembrane force probe that the damping time for the
system with tethered bonds is on the order of $10^{-3}$ s
\cite{a:NguyenKochMerkel2003}, which is typically at least one order
of magnitude smaller than the transition times for adhesion bonds
\cite{a:Merkel2001}. The polymer relaxation time is determined by the
Zimm time \cite{b:DoiEdwards1986}.  For a Flory chain, it is $\tau_Z =
\eta R_F^3 / \kt$ where $\eta$ is the viscosity of the surrounding
fluid, $R_F$ is the Flory radius $R_F \simeq a N^{0.6}$ with Kuhn
length $a$ and number of Kuhn segments $N$. The viscosity of water is
on the order of $\eta \sim 10^{-3} \rm Pa\, s$. For polymers with $R_F
\sim 10\ \rm nm$ and for $\kt \sim 4\ \rm pN\, nm$ one has $\tau_Z
\sim 10^{-7} \rm s$. This result agrees with earlier estimates for
polyethylene glycol chains
\cite{a:WongEtAl1997,a:MoreiraEtAl2003,c:moor06}.  The very fast
relaxation time scale for the polymers allows us to use a stationary
density distribution for the ligands.

\subsection{Dimensionless parameters}

The master equation \eq{MasterEquation} together with
\eqs{ReverseRate}{Force_i} for the reverse rate and
Eqs.~(\ref{ForwardRate}), (\ref{partitionsum}) and (\ref{density}) for
the forward rate completely specifies our model. For the following, it
is useful to introduce dimensionless quantities. First we introduce
dimensionless time $\tau = k_0 t$. Then we non-dimensionalize all
distances by writing them in units of the unstressed ligand tether
length $\ell_b$. The dimensionless relaxed ligand-receptor distance is
denoted by $\lambda = \ell / \ell_b$. We also introduce $\kappa = k_b
/ k_t$, the ratio of the spring constants of bonds and transducer, and
$\phi = k_b\ell_b/F_0$, the force in units of $F_0$ that is necessary
to extend a ligand tether spring by $\ell_b$, i.e.~to twice its
unstressed length. Then, the reverse rate reads
\begin{equation}\label{ReverseRate2}
r(i) = i \Exp{\phi\lambda/(1 + \kappa i)} = i \Exp{\phi\lambda(i)}\,,
\end{equation}
where $\lambda(i) = \lambda/(1 + \kappa i)$ has been introduced as an
abbreviation for the extension of the tethers (receptor-ligand distance
with $i$ closed bonds).  Regarding the association process, we define
the dimensionless on-rate $\gamma = (k_{on}/k_0) (\ell_{bind}/\ell_b)$,
which is weighted with the ratio of binding radius $\ell_{bind}$ and
unstressed tether length $\ell_b$, and the inverse thermal energy
$\kt$ non-dimensionalised by the tether energy at an extension equal to
their rest length, $\beta = k_b \ell_b^2 / 2 \kt$.  Then, the forward
rate reads
\begin{equation}\label{ForwardRate2}
g(i) = 2\gamma(N_t - i)\left[\frac{\beta}{\pi}\right]^{\12}
        \frac{\Exp{-\beta\lambda^2(i)}}
             {\Erf{\beta^{\12}} + \Erf{\beta^{\12}\lambda(i)}}\,.
\end{equation}
With the definition of these dimensionless rates, the dynamical
equations, that is master equation, mean field equation and
Fokker-Planck equation have the same form as in
Eqs.~(\ref{MasterEquation}), (\ref{DeterministicEquation}) and
(\ref{FokkerPlanckEquation}), but with time derivatives in $\tau$
rather than $t$.  Since reverse rate \eq{ReverseRate2} and forward
rate \eq{ForwardRate2} are both non-linear in $i$, the mean field
equation for $N$ is only valid for large system size.

\begin{table*}
\begin{center}
\begin{tabular}{|l|c|c|l|}
\hline\hline
definition & typical & integrins & meaning \\
\hline
$\beta    := k_b\ell_b^2/2 \kt$         & $0.1\dots10$ &  7    & inverse ligand temperature in units of tether energy\\
$\lambda  := \ell / \ell_b$             & $0.1\dots10$ &  0.75 & receptor-ligand distance \\
$N_t$                                   & $10 \dots 25$&  10   & cluster size\\
$\phi     := k_b\ell_b / F_0$           & $0.1$        &  0.27 & tether force in units of internal force scale of bonds \\
$\kappa   := k_b / k_t$                 & $1$          &  0.9  & ratio of tether and transducer stiffness\\
$\gamma   := \hat\gamma(\ell_{bind}/\ell_b)$ & $1$     &  1    & conditional single bond on-rate\\
\hline\hline
\end{tabular}
\caption{The six parameters of the model: definitions, typical values used
in the calculations, estimates for the integrin-fibronectin system and
meaning.}
\end{center}
\label{parameter}
\end{table*}

Our model now contains six dimensionless parameters. The number of
receptor-ligand pairs is given by the \textit{cluster size} $N_t$. The
\textit{conditional rebinding rate} $\gamma$ describes the rate of
binding with a flat density distribution (infinite temperature) on an
interval of length $\ell_b$. The \textit{relative stiffness of the
  tethers}, $\kappa = k_b / k_t$, implies the two limits of $\kappa
\to \infty$ (soft transducer) and $\kappa \to 0$ (stiff transducer).
In the following we will use the intermediate case $\kappa = 1$. The
\textit{dimensionless force constant} $\phi$ measures the force needed
to stretch the tethers to twice their unstressed length in units of
the intrinsic force scale of the adhesion bonds. For an entropic
spring, this essentially scales as the ratio of two length scales, the
bond reactive compliance $k_B T / F_0$ and the rest length of the
tethers. In practice it will have a rather small value and in the
following we use $\phi = 0.1$. The mobility of the ligands is
represented by the \textit{inverse ligand temperature} $\beta$. In the
limits of soft transducer ($\kappa \to \infty$) and very high ligand
temperature ($\beta \to 0$), our model simplifies to a case which we
have studied before in order to assess adhesion cluster stability
under force \cite{uss:erdm04a,uss:erdm04c}.  In this paper, we rather
focus on the role of ligand-receptor distance $\lambda$, which in
combination with the different spring constants replaces the
dimensionless force $f$ used in the earlier model. In experimental
setups, $\lambda$ is certainly the most accessible parameter. The
parameters and their definition are summarized in Tab.~I.
There, we also give the typical range of parameters which was used for
calculations and some estimates for the integrin-fibronectin system.

\section{Bifurcation analysis of the mean field equation}
\label{sec:bifurcation}

\FIG{TimeDerivative}{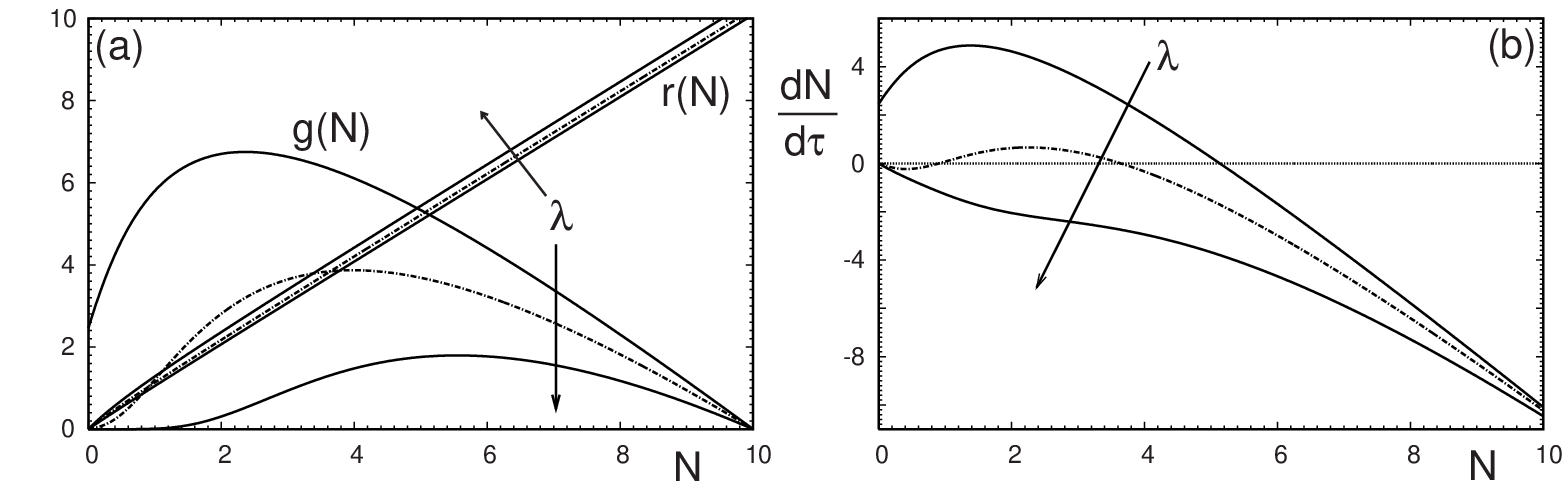}{(a) Reverse and forward rate $r(N)$ and $g(N)$ and (b)
the time derivative $dN/d\tau = g(N) - r(N)$ as a function of the number of closed bonds $N$ for
$N_t = 10$ and $\beta = 1$. The receptor-ligand distance is $\lambda = 1, 2.5$ and
$5$. The other parameters are $\gamma = 1$, $\kappa = 1$ and $\phi = 0.1$.}

To analyze the stationary solutions of the mean field equation, we
first consider the dependence of the reverse rate $r(N)$ from
\eq{ReverseRate2} and the forward rate $g(N)$ from \eq{ForwardRate2}
on the number of closed bonds $N$ and on the model parameters.
Stationary solutions are the fixed points of the
\eq{DeterministicEquation} and correspond to intersections of $r(N)$
and $g(N)$, because then the time derivative $dN/d\tau = g(N)-r(N)$
vanishes.  \fig{TimeDerivative}(a) plots $r(N)$ and $g(N)$ as function
of $N$ for different values of the relaxed receptor-ligand distance
$\lambda$.  The single bond off-rate, $k_{off}(N) = \Exp{\phi \lambda
  / (1 + \kappa N)}$, is finite at $N=0$, thus $r(0) = 0$. With
increasing $N$, the reverse rate $r(N)$ increases almost linearly,
although the single bond off-rate is a monotonous decreasing function
of $N$. The weak influence of the exponential off-rate $k_{off}(N)$ is
mainly due to the small force constant $\phi = 0.1$: the entropic
tether force does not suffice to accelerate bond rupture appreciably.
For larger $\phi$ the reverse rate increases quickly from $r(0) = 0$
and can have a local maximum and minimum at small $N$. For large $N$
the linear term dominates in any case. Alternatively, non-monotonous
behavior could be induced by a large $\kappa$, that is for a soft
transducer. The forward rate $g(N)$ vanishes for $N = N_t$, goes
through a maximum at intermediate $N$ and then decreases. At $N=0$ the
forward rates is always positive and approaches zero only in the limit
of infinite $\beta$ or $\lambda$. At intermediate values for
$\lambda$, there are three intersections of $g(N)$ and $r(N)$.
\fig{TimeDerivative}(b) plots $dN/d\tau$ as function of $N$ for the
same set of parameters as in \fig{TimeDerivative}(a). It is positive
at $N=0$, because $r(0) = 0$ and $g(0) > 0$. At intermediate $N$ and
small $\lambda$, the time derivative has a maximum which reflects the
maximum of $g(N)$ before it becomes negative at large $N$ where the
reverse rate dominates. The fixed point at large $N$ is stable and
represents the bound state of the adhesion cluster containing a large
number of closed bonds.  For intermediate $\lambda$ there are three
fixed points, including two stable fixed points at large $N$ (bound
state) and $N\approx 0$ (unbound state). The two stable fixed points
are separated by an unstable one. At large $\lambda$ there is only one
stable, unbound state which approaches $N=0$ in the limit of large
$\lambda$.

\FIG{bifurcation}{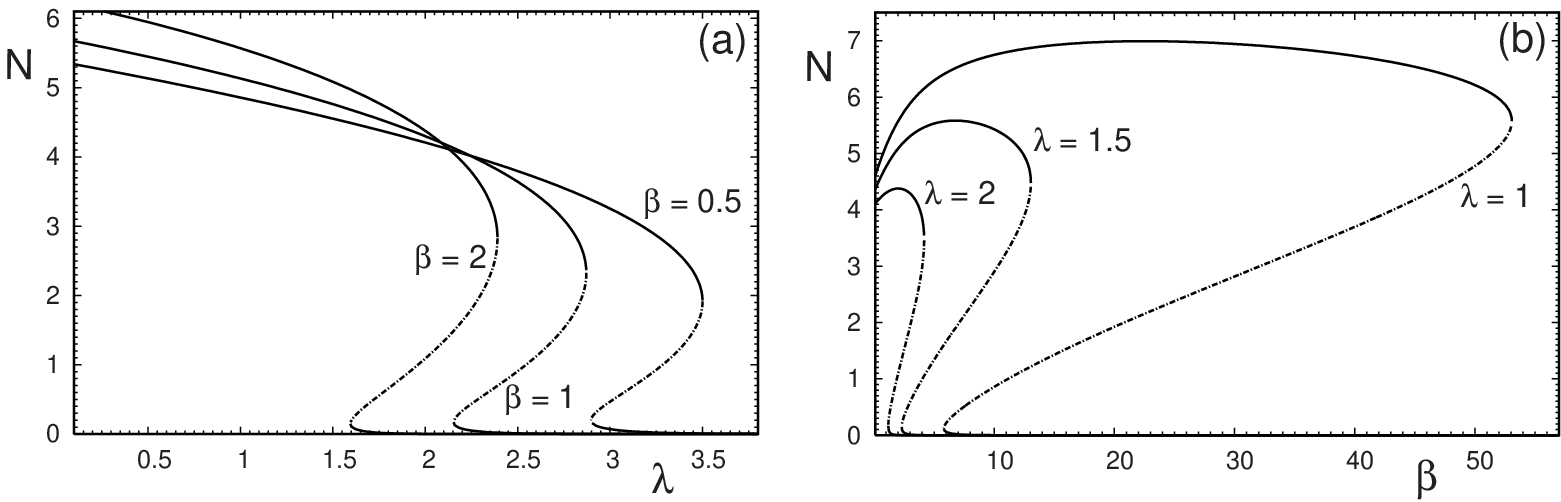}{One-parameter bifurcation diagrams showing the fixed
points of the mean field equation for the cluster size $N_t = 10$ as function
of (a) $\lambda$ for $\beta = 0.5, 1$ and $2$ and (b) $\beta$ for $\lambda = 1,
1.5$ and $2$. The stable stationary states are the solid lines, the unstable fixed
points are dash-dotted. The other parameters are $\gamma = 1$, $\kappa = 1$ and
$\phi = 0.1$.}

\fig{bifurcation} summarizes the behavior of the fixed points in the
form of two one-parameter bifurcation diagrams which show the fixed
points as function of $\lambda$ and $\beta$, respectively. At small
separation, a single stable fixed point exists at large $N$. Here, the
adhesion cluster is bound because the force on the bonds is small and
the density of free ligands close to the receptor is large, therefore
rupture events are rare and can be balanced by rebinding. With
increasing $\lambda$, the force on the bonds increases and the number
of closed bonds in the bound state decreases. At large $\lambda$,
there is a single stable fixed point at $N\approx 0$. Here, the
adhesion cluster is unbound because forces are large and ligand
density at the receptors is small so that rupture events occur
frequently and cannot be balanced by rebinding. The transition from
bound to unbound proceeds via two saddle-node bifurcations. At small
$\lambda$, the stable unbound fixed point appears together with an
unstable fixed point separating the stable ones. The unstable fixed
point merges with the bound stable fixed point at larger $\lambda$. In
the window of bistability between the two bifurcations, two stable
fixed points coexist. The position of this window moves to smaller
$\lambda$ with increasing $\beta$. The bifuraction behavior as
function of the inverse ligand temperature $\beta$ is qualitatively
similar although the position of the stable bound state initially
increases with $\beta$ because the forward rate initially increases.
With decreasing $\lambda$, the position of the bistable range shifts
to larger $\beta$ while its width increases strongly.

\FIG{PhaseDiagram}{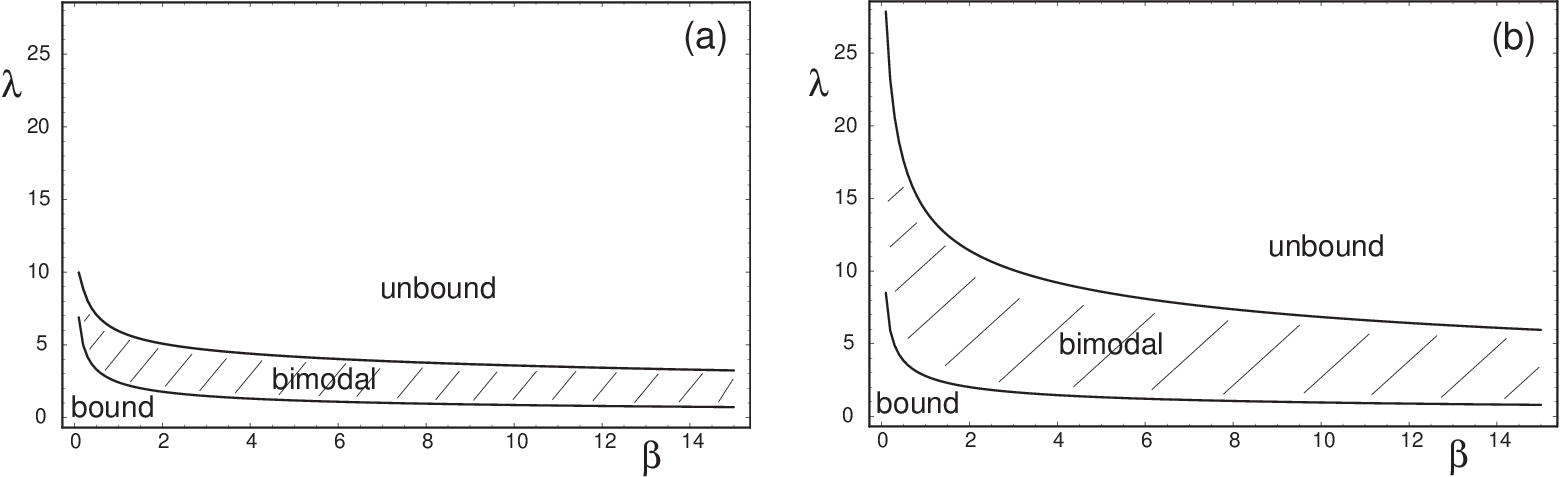}{Stability diagram for adhesion clusters: solid lines are
numerically determined positions of the lower and upper bifurcations as function
of $\beta$ and $\lambda$. The shaded area between the curves is the region of
bistability, above this region there is a single, unbound state while below it a
single stable bound cluster exists. The curves are derived for the cluster size
$N_t = 10$, $\kappa = 1$, $\phi = 0.1$ and (a) $\gamma=1$ and (b) $\gamma=5$.} 

In \fig{PhaseDiagram} we construct a stability diagram which
identifies the region of bistability as a function of $\lambda$ and
$\beta$. The positions of the bifurcation delineating the bistable
region were determined numerically. For large $\beta$ the width of the
interval in $\lambda$ stays almost constant while the position
decreases slowly which explains the very large bistable range in
$\beta$ at small $\lambda$. In general, $\lambda$ and $\beta$ are
inversely related. For the lower bifurcation one can use the
approximate criterion that the slope of $dN/d\tau$ with respect to $N$
has to become negative.  Neglecting $r'(N)$, one is left with the
condition $dg(N)/dN = 0$ at $N=0$. For large $\beta \gg 1$ one finds
$\lambda^2 \simeq 1/(2\beta\kappa)$.  For $\beta \ll 1$ and for
sufficiently large $\lambda$, on the other hand, one finds $\beta
\simeq \ln(\kappa N_t)/\lambda^2$. Thus in both limits, $\beta$ and
$\lambda$ are related by an inverse square root.  At small $\beta$,
the width of the bistable range decreases as the two curves eventually
converge and bistability vanishes. The stability diagram can be
regarded as the projection of the cusp-like surface of the fixed
points on the $(\lambda,\beta)$ plane. For large clusters and also for
larger on-rates $\gamma$, the curves will meet at negative $\beta$ and
bistability persist for positive $\beta$.

In the framework of biochemical control of biological systems,
bistability is commonly associated with an underlying positive
feedback mechanism \cite{sb:TysonChenNovak2003}. In our case,
bistability can arise from two positive feedback mechanisms as
described above. First, there is positive feedback for bond rupture:
as one bond breaks, the force on the remaining bonds increases, thus
increasing their dissociation rate. Second, there is positive feedback
for binding: as one ligand binds a receptor, the receptor-ligand
distance is decreased and the binding rate for the other ligands is
increased. In general, we verified that in our model, both mechanisms
can lead to bistability. However, for the parameter range chosen here
it is only the positive feedback of binding which is responsible for
the observed bistability. As shown in \fig{TimeDerivative}, the
reverse rate $r(N)$ increases almost linearly for the set of
parameters chosen. The forward rate $g(N)$, on the other hand, is
non-monotonous. Thus for the parameter range chosen here, the positive
feedback underlying bistability is mostly due to the forward rate
$g(N)$.

\section{Bifurcation analysis with a stochastic potential}

\subsection{Stationary solution of the Fokker-Planck equation}

The Fokker-Planck equation \eq{FokkerPlanckEquation} has the stationary solution
\begin{equation}\label{StationaryDistribution}
P^s(i) = \frac{C}{D(i)}\Exp{2\int_0^i\frac{A(i')}{D(i')}di'}\,,
\end{equation}
where $C$ is a normalization constant. The integrand in the exponent exists and the
expression is integrable because the Fokker-Planck coefficients are finite and
defined on a compact interval. In the absence of sources and sinks the flux
\begin{equation}
J^s(i) = A(i)P^s(i) - \frac{1}{2}\left\{D'(i)P^s(i) +  D(i)P^s(i)'\right\}
\end{equation}
has to vanish. The derivative of $P^s(i)$ with respect to $i$ is
\begin{equation}\label{FirstDerivative}
P^s(i)' = \frac{2 A(i) - D'(i)}{D(i)}P^s(i)\,,
\end{equation}
so that indeed
\begin{equation}
J^s(i) = \left\{A(i) - \frac{1}{2}\left\{D'(i) + 2 A(i) - D'(i)\right\}\right\}P^s(i) = 0\,. 
\end{equation}
In particular, the flux through the boundaries vanishes, as required for
reflecting boundaries.

\subsection{Stochastic potential}

The stationary probability distribution \eq{StationaryDistribution}
can be used to define an energy landscape $E(i)$ by
\begin{equation}
E(i) = -\log{P^s(i)}\,.
\end{equation}
The extrema of this potential are determined by the condition
\begin{equation}\label{StochasticPotential}
\frac{dE(i)}{di} = -\frac{1}{P^s(i)}\frac{dP^s(i)}{di} = 0
\quad\Leftrightarrow\quad \frac{dP^s(i)}{di} = 0 \,.
\end{equation}
With \eq{FirstDerivative} for the first derivative of the
distribution, the position of the extrema of the stochastic potential
$E(i)$ are thus determined by
\begin{equation}
A(i) - \frac{1}{2}D'(i) = 0 \,.
\end{equation}
These extrema have a similar physical meaning as the fixed points of
the mean field equation, but they are more rigorous in including
thermal fluctuations. In the framework of the stochastic potential,
bistability requires a bimodal potential landscape in which two minima
of the stochastic potential coexist. The coexisting minima (maxima of
the probability distribution) are separated by a potential barrier
(minimum of the probability distribution). The separated peaks of the
probability distribution in these minima are commonly referred to as
\textit{macrostates} of the stochastic system because they are
possible realizations of a macroscopic, deterministic system. One can
regard the extrema as the fixed points of the dynamical system
\begin{equation}\label{StochasticEquation}
\frac{di}{dt} =  A(i) - \frac{1}{2}D'(i) = g(i)-r(i) - \frac{1}{2}(g'(i)+r'(i))\,.
\end{equation}
This is the mean field equation corrected for the effects of
non-homogeneous mobility.  \eq{StochasticEquation} allows to determine
extrema in the same way as the fixed points of the mean field equation
in the previous section. For a constant diffusion coefficient the
fixed points are identical to the extrema of the stochastic potential.
Non-homogeneous diffusion terms change the position of fixed point and
can even destroy fixed points or create new ones (\emph{noise-induced
  transitions}) \cite{b:hors84}.

\subsection{Bifurcation analysis}

\FIG{bifex}{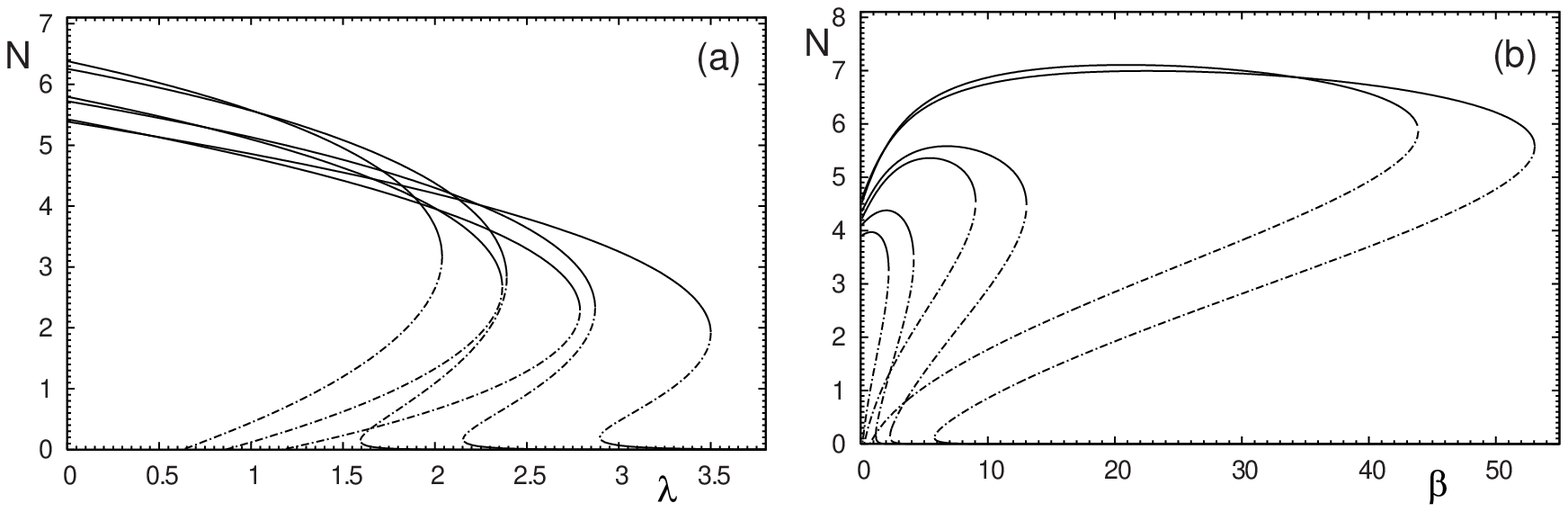}{Comparison of the extrema of the
  stochastic potential with the deterministic fixed points as function
  of (a) $\lambda$ with $\beta = 0.5$, $1$ and $2$ and (b) $\beta$
  with $\lambda = 1$, $1.5$ and $2$ for the cluster size $N_t = 10$.
  For small parameters $\lambda$ and $\beta$, stochastic and deterministic
  results agree well for the stable fixpoints at large $N$, but differ
  strongly in the transition regions.}

Using the stochastic dynamic system, the topology of the stochastic
potential can be analyzed just as the mean field equation by deriving
bifurcation diagrams showing the positions of the macrostates (the
extrema of the potential) as function of the model parameters. The
extrema are determined as stable and unstable fixed points of
\eq{StochasticEquation}. \fig{bifex} shows these stochastic
bifurcation diagrams as function of receptor-ligand distance $\lambda$
in comparison to those from the mean field equation. For sufficiently
small $\lambda$, the upper stable fixed point of the deterministic
equation agrees well with a bound macrostate at large $N$ in the
stochastic potential. The saddle-node bifurcation in which the bound
macrostate vanishes occurs at much smaller separation $\lambda$ than
in the deterministic picture, thus fluctuations destabilize the
adhesion cluster. The minimum of the stochastic potential lies above
the unstable deterministic fixed point and becomes negative at small
$\lambda$. For positive $i$, the stochastic potential has no second
minimum, but it has a boundary minimum at $i=0$. This second
macrostate corresponds to the unbound fixed point of the mean field
equation.  For larger clusters, the agreement between the fixed points
of the stochastic potential and the mean field equation improves.  In
the limit of very large clusters, the two solution approach each other
as shown in \fig{bifex_large} for $N_t = 100$ and $250$. In this case,
the unbound state is practically indistinguishable from $i=0$.

\FIG{bifex_large}{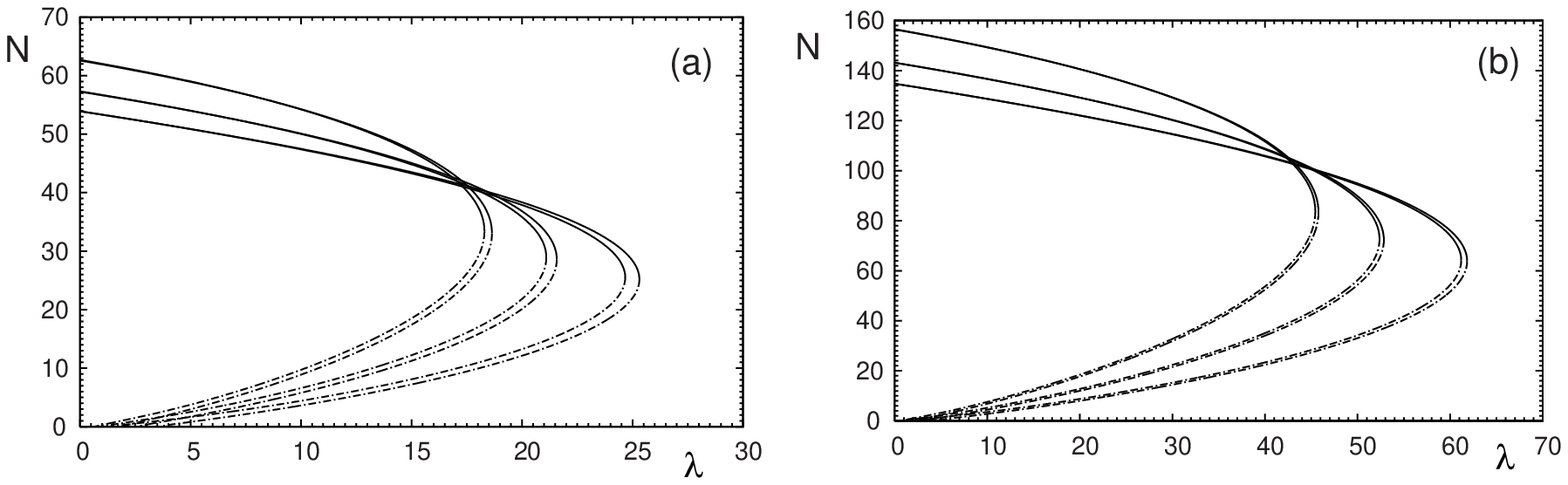}{Extrema of the stochastic
  potential as function of $\lambda$ compared with the deterministic
  fixed points for the same parameters as in \fig{bifex}, but for
  cluster sizes (a) $N_t = 100$ and (b) $N_t = 250$. For these
  large system sizes, stochastic and deterministic results agree well
  over the full range of parameters.}

\section{Stationary solutions of the master equation}
\label{sec:meqstat}

For a one-step master equation on a finite range without sources or
sinks, stationarity $\dot p_i(\infty) = 0$ implies detailed balance,
that is $r(i)p_i(\infty) = g(i-1) p_{i-1}(\infty)$. Iterating this
relation results in the stationary probability distribution
\begin{equation}\label{StatDist}
\frac{p_i(\infty)}{p_0(\infty)} = \frac{g(0)}{r(i)}\prod_{j=1}^{i-1}\frac{g(j)}{r(j)}\quad\text{for $0 < i \le N_t$} \,.
\end{equation}
The normalization constant is the stationary state probability for the completely
dissociated state $i=0$,
\begin{equation}
p_0(\infty) = \left( 1 + \sum_{i=1}^{N_t}\frac{g(0)}{r(i)}\prod_{j=1}^{i-1} \frac{g(j)}{r(j)} \right)^{-1} \,.
\end{equation}
\fig{StatDist} shows a density plot of the stationary distribution
$\left\{p_i(\infty)\right\}_{i=0}^{N_t}$ as function of relaxed
receptor-ligand distance $\lambda$ for cluster sizes $N_t=10$ and
$N_t=25$. For small $\lambda$, there is a single peak at a finite
number of closed bonds which is broadened by fluctuations. This
corresponds to the bound state of adhesion clusters.  For large
$\lambda$, there is a single maximum at the completely dissociated
state $i=0$, which is the unbound state. In an intermediate range of
$\lambda$, the stationary distribution has two maxima and bound and
unbound adhesion clusters coexist. Thus the full stochastic model
indeed shows bistability as suggested by the bifurcation analysis of
the mean field equations. For the smaller cluster size $N_t = 10$,
fluctuations are large and the transition from bound to unbound
appears rather smooth. For larger systems, the transition becomes
sharper and discontinuous. This discontinuity is demonstrated by the
average number of closed bonds in the adhesion cluster which shows a
steep decrease as the occupancy switches from bound to unbound.  For a
bimodal distribution with two distinct macrostates, average numbers of
closed bonds can be defined in the two peaks separately. We use the
probability functions $\left\{p_i(\infty)\right\}_{i=3}^{N_t}$ for the
upper and $\left\{p_i(\infty)\right\}_{i=0}^{1}$ for the lower peak
with proper normalization to calculate the average number of closed
bonds in the two macrostates. \fig{StatDist} shows that these averages
vary slightly with $\lambda$ and the steep decrease of the full
average is mostly due to the change in occupancy probability than in
the position of the peaks. \fig{StatDist} also shows the bifurcation
result from the mean field equation. For the larger system $N_t = 25$,
the position of the maxima are in good agreement with the fixed points
and the onset of bistability at small $\lambda$ agrees well with the
first bifurcation. For the smaller system $N_t = 10$, the agreement
between fixed points and maxima is still good, but the onset of
bimodality is overestimated by the lower bifurcation.  Here, the
stochastic potential yields a much better estimate for the locations
of the bifurcation (not shown). For growing cluster size, the range of
parameters in which the coexisting macrostates are occupied to a
similar degree shrinks, although the range of bistability as revealed
by the deterministic equation or the stochastic potential grows.  In
analogy to first order phase transitions, the discontinuous transition
from bound to unbound will occur at a sharp parameter value in the
limit of infinite clusters.

\FIG{StatDist}{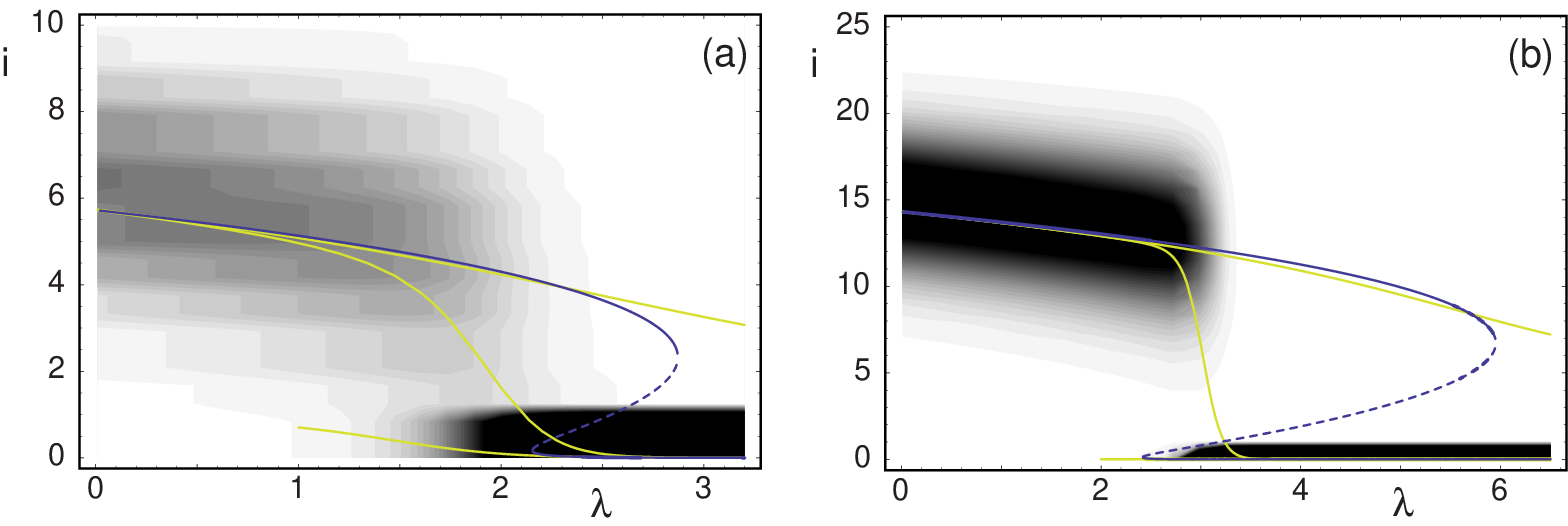}{Density plot of the stationary probability
  distribution of the master equation $\{p_i(\infty)\}_{i=0}^{\infty}$
  as function of separation $\lambda$ for cluster sizes (a) $N_t = 10$
  and (b) $N_t = 25$.  The other parameters are $\beta = 1.0$, $\gamma
  = 1$, $\kappa = 1$ and $\phi = 0.1$.  Dark regions indicate high
  probability.  The curves are the average number of closed bonds in
  the bound and unbound macrostates, in the full distribution and as
  predicted from the mean field equation, respectively.}

Biological systems are usually of finite size and time dependent
processes are often important. Whether coexisting states and
transitions between them can be observed depends on the relative
occupancy of the coexisting states and on the time scale for
transitions between the states. The bare presence of bistability
identified in the deterministic equation is irrelevant if the time
scale for the transitions is larger than the typical observation
times. The time scale for transitions will also be important if
parameters are time dependent. If the change of parameters is faster
than the time scale for equilibration of the probability distribution,
metastable states will be populated and hysteresis in the transition
parameters will be observed. In the following section, we therefore
analyze dynamic properties of the master equation.

\section{Dynamic properties of the master equation}\label{sec:meqdyn}

\FIG{trajectory}{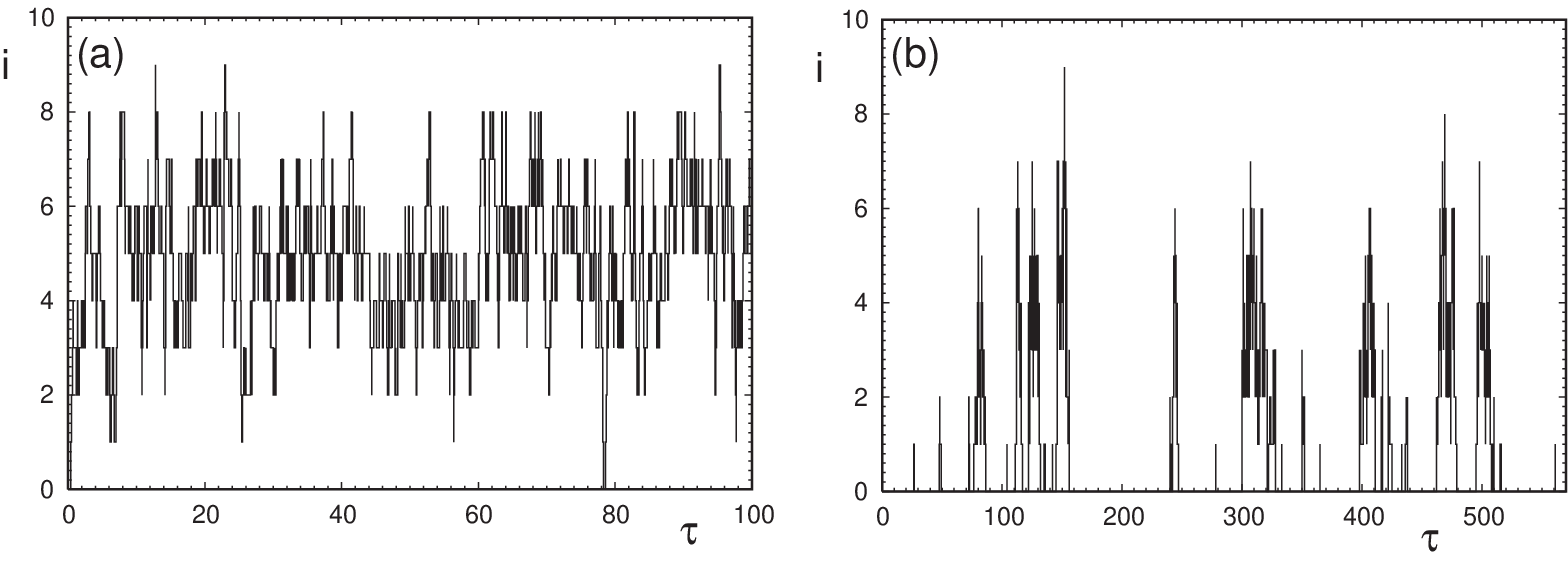}{Single adhesion cluster trajectories for the
cluster size $N_t = 10$ at $\beta = 0.5$ at two different values of (a)
$\lambda = 1$ and (b) $\lambda = 3$ generated with the Gillespie
algorithm. Averaging over time or over many trajectories gives the
stationary probability distribution.}

The stationary probability distribution \eq{StatDist} arises by
averaging over many individual trajectories or over a single
trajectory for a long time.  \fig{trajectory} shows two sample
trajectories for $N_t = 10$ bonds and $\beta = 0.5$ at $\lambda = 1$
and $\lambda = 3$, respectively. The stochastic trajectories are
generated with the Gillespie algorithm for exact stochastic
simulations \cite{mc:Gillespie1976,mc:Gillespie1977} and show how the
number of closed bonds $i$ changes over time. For $\lambda = 1$, the
stationary distribution is unimodal and has a single peak around the
bound state. Due to the small cluster size, the number of closed bonds
fluctuates strongly around the average; occasionally it reaches the
completely dissociated state, but rebinding takes place immediately.
This leads to a single, broad peak as in \fig{StatDist} at small
$\lambda$. For $\lambda = 3$ the stationary distribution is bimodal.
The sample trajectory alternates between bound and unbound state. As
long as the time spent in the respective states is large enough, the
time taken for the actual transition is negligible. When bound, the
trajectory fluctuates around an average as in (a). An encounter of the
completely dissociated state, however, is usually followed by a longer
time with no or few bonds.  Single closed bonds are formed
occasionally, but this rarely leads to the formation of a large number
of closed bonds. Increasing the receptor-ligand distance further
increases the time spent in the unbound state relative to that in the
bound state.  This means that the occupancy probability of the bound
state is reduced. The trajectories alternating between the states with
very short transition times yield the bimodal stationary
distribution. The probability to find the system in one of the states
is proportional to the time spent in that state before a transition.
For very large $\lambda$ with unimodal distribution, binding of
trajectories does not take place with appreciable frequency.

The dynamic properties of the stochastic model can be characterized by
the mean first passage time $T_{m,n}$ between two states $m$ and $n$,
that is, the time it takes on average to reach $n$ for the first time
from $m$. To elucidate the relevance of the fixed points in the
stochastic system, the transition times from the unbound state $m=0$
to the bound state $n \simeq N_t/2$ and vice versa are of particular
interest. The value $N_t/2$ can be used because in this context it is
only relevant that the bound state is above the transition barrier
between the two macrostates; the dynamics within the respective
bassins of attraction are much faster than the dynamics across the
barrier. For a one-step master equation like \eq{MasterEquation}, the
mean first passage time from the initial state $m$ to the final state
$n$ satisfies the recursion relation \cite{b:VanKampen2003}
\begin{equation}
g(n)\left\{T_{m+1,n} - T_{n,m}\right\} + r(m) \left\{T_{m-1,n} - T_{m,n}\right\} = -1
\end{equation}
with the boundary condition $T_{m,m}=0$. For a transition from a state $m$ to $n >
m$, that is for an increase of the number of closed bonds,
one has
\begin{equation}\label{mfpt_up}
T_{m,n} = \sum_{i=m}^{n-1} \left\{\frac{1}{g(i)}
        + \sum_{j=0}^{i-1}\frac{1}{g(j)}\prod_{k=j+1}^{i}\frac{r(k)}{g(k)}\right\}\,.
\end{equation}
For the reverse transition from a state $m$ to $n < m$, where the number of closed
bonds decreases, one has
\begin{equation}\label{mfpt_down}
T_{m,n} = \sum_{i=n+1}^{m} \left\{\frac{1}{r(i)}
        + \sum_{j=i+1}^{N_t}\frac{1}{r(j)}\prod_{k=i}^{j-1}\frac{g(k)}{r(k)}\right\}\,.
\end{equation}
The first term in curly brackets in \eqs{mfpt_up}{mfpt_down} is the mean
first passage time for a trajectory exclusively with binding or rupture,
respectively.  The second term with the product over the ratio of
rebinding and rupture rates describes the increase of the mean first
passage time through backward reactions, that is rupture if $m < n$ and
rebinding if $m > n$.

\FIG{mfpt}{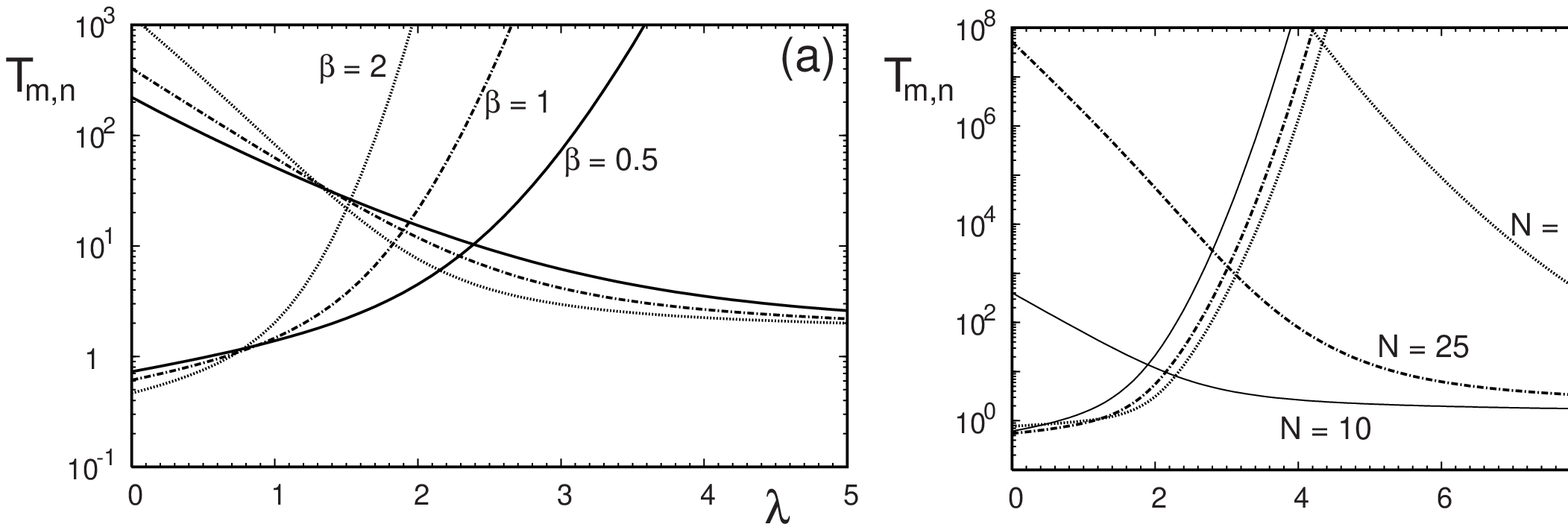}{Mean first passage times $T_{m,n}$ from
\eqs{mfpt_up}{mfpt_down} for transitions between unbound state $m=0$ and
unbound state $n=N_t/2$ as function of $\lambda$ for (a) $N_t = 10$ and
$\beta = 0.5, 1$ and $2$ and (b) for $\beta = 1$ with $N_t = 10, 25$ and
$50$. The other parameters are $\gamma = 1$, $\kappa = 1$ and $\phi =
0.1$.}

\fig{mfpt} plots the mean first passage times $T_{m,n}$ from
\eqs{mfpt_up}{mfpt_down} for transitions from the unbound state $m=0$
to the bound state $n=N_t/2$ (binding time) and from the bound state
to the unbound (unbinding time). For parameter values for which only a
single stable macrostate exists, the transition time into this state
is very small and on the order of magnitude of transitions between
neighboring states. The time for transitions in the reverse direction
becomes extremely large for large clusters. The range of bistability
is characterized by binding and unbinding times which are both larger
than the single step transition times. Because barrier crossing itself
is a fast process, the ratio of binding to unbinding time equals the
ratio of occupancy of the two macrostates in the range of bistability.
The point where the transition times are equal thus defines a
stochastic transition point at which both states are equally occupied.
The plots of the transition times as function of $\lambda$ for
different $\beta$ at $N_t = 10$ in \fig{mfpt}(a) show that this
transition point shifts to larger $\lambda$ with decreasing $\beta$,
because with increasing temperature (ligand mobility), larger
separations can be bridged by the ligand tethers.  \fig{mfpt}(b) plots
the transition times for $\beta=1$ at different $N_t$. This shows that
the stochastic transition point shifts to larger separation values
with growing cluster size.  At the same time the corresponding
switching times grow super-exponentially fast. This implies that for a
given separation and growing cluster size, the bound macrostate will
effectively become the only stable state; frequent switching between
bound and unbound states can thus only occur for small clusters. It is
important to note that these important conclusions can only be drawn
by considering the full stochastic dynamics.

Stochasticity of binding and strong dependence of the mean first
passage time on $\lambda$ has to be considered in measurements of the
binding ranges of tethered ligands. If the receptor-ligand distance is
reduced step by step and the transducer is held at a constant distance
during a short period of time $\tau_s$, binding will typically be
observed at a distance where the mean first passage time from unbound
to bound $T(\lambda)$ is comparable to $\tau_s$. In general, binding
is always possible for all distances smaller than the contour length
of the tethers. The actual binding distance is a stochastic variable.
If transitions proceed with the constant rate $1/T(\lambda)$, the
probability to observe binding after the $n^{th}$ step at a distance
$\lambda_n$ is
\begin{equation}
p = \left(1 - \Exp{-\tau_s / T(\lambda_n)}\right)\prod_{i=1}^{n-1} \Exp{-\tau_s / T(\lambda_i)}\,.
\end{equation}
With the strong dependence of the binding time on $\lambda$ as shown
in \fig{mfpt} this distribution will be have a sharp peak when
$T(\lambda_n) \leq \tau_s \leq T(\lambda_{i < n})$. The
probability to bind at any distance below a given $\lambda_n$ is then
close to a step function as observed experimentally and theoretically
\cite{a:JeppesenEtAl2001,a:MoreiraEtAl2003}.

\section{Extension to worm-like chain model}\label{sec:wlc}

Unlike in the simple harmonic spring model, real polymers are not
infinitely extensible, but are characterized by a finite contour
length. A commonly used model for real polymers is the so-called
worm-like chain or Kratky-Porod model \cite{b:DoiEdwards1986}. It has
been used before to model semi-flexible biopolymers like DNA, F-actin
or titin \cite{a:MarkoSiggia1995}.  A worm-like chain is characterized
by the contour length $L$ and the persistence length $L_p$ which
describes the bending stiffness of the filament. The worm-like chain
model can also be extended to include elasticity of the polymer
backbone which allows stretching beyond the contour length
\cite{a:KierfeldEtAl2004}. The forces needed to stretch the polymer
monomers are much larger than the typical thermal forces and will not
be considered in the following.

\subsection{Force extension relation and rupture rate}

For a worm-like chain, the force $F_{wlc}$ which induces an average
extension $x$ of the worm-like chain polymer can be approximated by
the interpolation formula \cite{a:MarkoSiggia1995}
\begin{equation}\label{wlcforce}
F_{wlc}(x) = \frac{\kt}{L_p}\left\{\frac{1}{4(1-(x/L))^2}+\frac{x}{L}-\frac{1}{4}\right\}\,.
\end{equation}
With the first term in curly brackets the force diverges as $x$
approaches the contour length $L$. The second term describes harmonic
behavior at small extensions with a force constant $3 \kt / 2 L L_p$.
The third, constant term guarantees that the force vanishes at
vanishing extension. We express \eq{wlcforce} in non-dimensional units
by writing extension in units of the contour length, $\xi = x / L$,
and force in units of the intrinsic force scale $F_0$ of adhesion
bonds, $f_{wlc} = F_{wlc} / F_0$. The force extension relation
\eq{wlcforce} then reads
\begin{equation}
f_{wlc}(\xi) = \phi_{wlc}\left\{\frac{1}{4(1 - \xi)^2}+ \xi - \frac{1}{4}\right\}
\end{equation}
where the ratio $\phi_{wlc}=\kt/(F_0 L_p)$ is defined in analogy to
$\phi$.  For small extension, $\xi\ll 1$, the constant
of proportionality between $\xi$ and $f_{wlc}$ is $3 \phi / 2$.

The extension $\xi_b(i)$ of bound tethers is determined by mechanical
equilibrium between tethers and transducer. In non-dimensional units
this reads
\begin{equation}\label{MechEquilWlc}
\frac{2 i\kappa_{wlc}}{3} \left\{\frac{1}{4(1-\xi_b(i))^2}+\xi_b(i)
-\frac{1}{4}\right\} = \lambda_{wlc}-\xi_b(i)
\end{equation}
which has to be solved for $\xi_b(i)$. The parameter $\lambda_{wlc} =
\ell / L$ is the non-dimensional relaxed receptor-ligand distance and
$\kappa_{wlc} = (3 \kt / 2 L L_p) / k_t$ measures the ratio of the
harmonic force constant of the polymer and the force constant of the
transducer. The two parameters are analogous to $\lambda$ and $\kappa$
for the spring model. Solving \eq{MechEquilWlc} for $\xi_b(i)$ yields
tether extension and force $f_{wlc}(\xi_b(i))$ as function of the
number of bound tethers alone. This result has to be inserted into
Bell's expression $k_{off} = \e{f_{wlc}(i)}$, leading to the reverse
rate of the adhesion cluster
\begin{equation}\label{WLCRuptureRate}
r(i) = i \e{f_{wlc}(i)}
\end{equation}
which has to be used in the one-step master equation
\eq{MasterEquation} and the mean field equation
\eq{DeterministicEquation}. As a polynomial of third order,
\eq{MechEquilWlc} can be easily solved for $\xi_b(i)$.

\subsection{Rebinding rate}

The energy needed to stretch a worm-like chain from zero extension to an extension
$x_b$ is given by the integral over the force $F_{wlc}$
\begin{equation}
V_{wlc}(x) = \int_0^{x_b(i)}F_{wlc}(x')dx'\,.
\end{equation}
In non-dimensional units the energy is
\begin{equation}
v_{wlc}(\xi_b(i)) = \frac{1}{\Lambda_p}\left\{\frac{1}{4(1-\xi_b(i))}+\frac{\xi_b^2(i)}{2}-\frac{\xi_b(i)}{4}-\frac{1}{4}\right\}
\end{equation}
where the dimensionless persistence length is $\Lambda_p = L_p / L$. With the
Boltzmann factor $\e{-v_{wlc}(\xi_b)}$ the density of unbound ligands at the
transducer is
\begin{equation}
\rho(\xi_b(i)) = \e{-v_{wlc}(\xi_b(i)))}/Z(\xi_b(i))
\end{equation}
where
\begin{equation}\label{wlcPartitionSum}
Z(\xi_b(i)) = \int_0^1 \e{-v(\xi)}d\xi + \int_0^{\xi_b(i)}\e{-v_{wlc}(\xi)}d\xi
\end{equation}
is used as the partition sum. The first term is added to prevent the
density from diverging at $\xi_b = 0$; for $v(\xi)$ a harmonic
potential with the same force constant as for the worm-like chain was
used. It takes into account that for entropic reasons the ligands are
found on average at a certain height above the substrate. The exact
distribution of the polymers in the presence of a wall is unknown
\cite{a:KochSommerBlumen1996}. In \cite{a:JeppesenEtAl2001}, it has
been calculated by Monte Carlo simulations for a bead-pearl model. It
was found that at large extensions, the distribution resembled that of
a freely jointed chain \cite{b:DoiEdwards1986}. The exact form of the
distribution should have no influence on the generic aspects we are
interested in. The distribution below the rest length is used only for
normalization but is irrelevant for binding.

The forward rate of the adhesion cluster as function of the number of closed bonds
$i$ is
\begin{equation}\label{wlcRebindingRate}
g(i) = \gamma(N_t - i)\rho(i) \,,
\end{equation}
where $\gamma = (k_{on}/k_0)(\ell_{bind}/L)$ has been defined.
\eqs{WLCRuptureRate}{wlcRebindingRate} together with the master equation
\eq{MasterEquation} and the deterministic equation \eq{DeterministicEquation}
define the dynamical system. The six dimensionless parameters for the worm-like
chain model are defined in analogy to those for the harmonic model. In the
following we use $\Lambda_p = 1$, $\gamma = 1$, $\kappa_{wlc} = 1.5$ and $\phi_{wlc}
= 0.1$. The choices for $\kappa_{wlc}$ and $\phi_{wlc}$ mean that the properties
of the tethers at small extension are similar as above for the harmonic tethers.

\subsection{Analysis of the worm-like chain}

\FIG{WLCpstat}{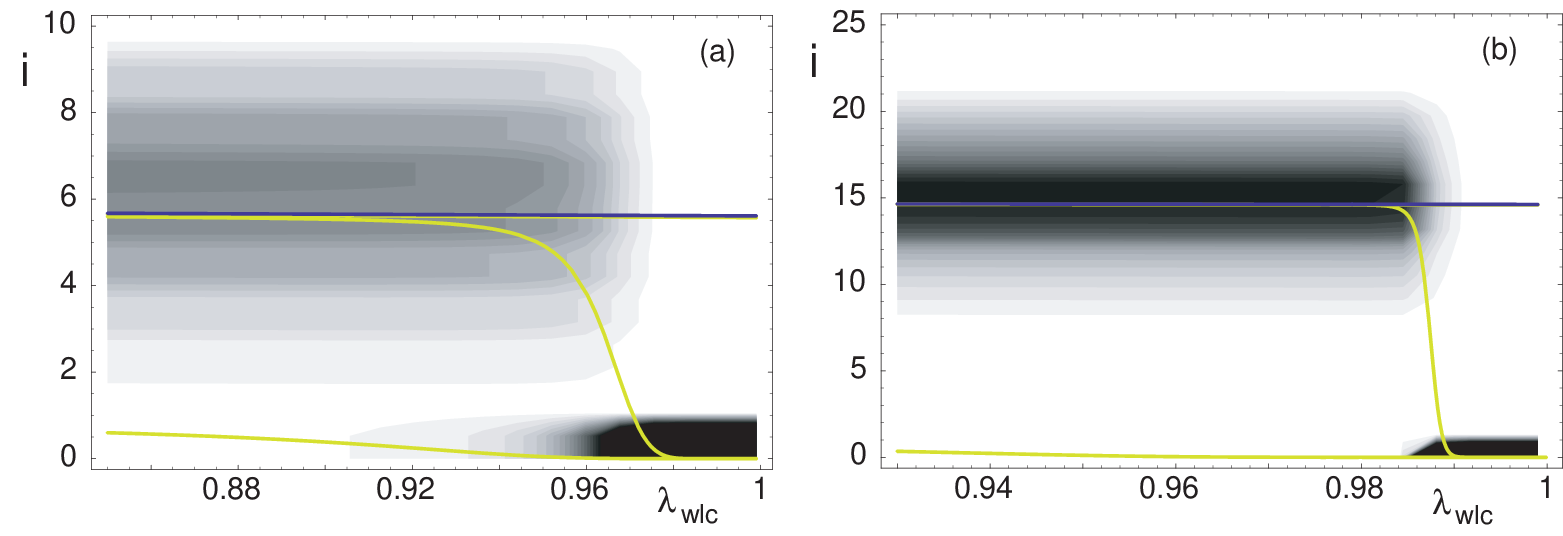}{Stationary probability distribution
  $\{p_i(\infty)\}_{i=0}^{N_t}$ from \eq{StatDist} with the transition
  rates \eq{WLCRuptureRate} and \eq{wlcRebindingRate} for the
  worm-like chain model and for cluster sizes (a) $N_t = 10$ and (b)
  $N_t = 25$ plotted as a function of the relaxed receptor-ligand
  distance $\lambda_{wlc}$. The other parameters are $\kappa_{wlc} =
  1.5, \phi_{wlc} = 0.1, \Lambda_p = 1$ and $\gamma = 1$. The curves
  show the upper stable fixed point of the mean field equation, the
  average number of closed bonds in the bound and the unbound
  macrostate, and in the whole cluster, respectively.}

A steady state analysis of the mean field equation
\eq{DeterministicEquation} with the transition rates
\eq{WLCRuptureRate} and \eq{wlcRebindingRate} from the worm-like chain
model shows that the bifurcation behaviour is very similar to the one
obtained for the linear springs, except that distances larger than
$\lambda_{wlc} = 1$ are not possible due to the finite contour length.
\fig{WLCpstat} shows a density plot of the stationary probability
distribution \eq{StatDist} for the worm-like chain transition rates
\eq{WLCRuptureRate} and \eq{wlcRebindingRate} as function of
$\lambda_{wlc}$. Together with the distribution, the dependence of the
upper stable fixed point of the mean field equation and the average
number of closed bonds of the full distribution as well as that of the
bound and unbound macro-states are displayed. Again the binding region
is bounded by the maximum extension $\lambda_{wlc} = \ell/L = 1$. As
for the harmonic tethers a bimodal region is found in which two
macrostates coexist. The average number of closed bonds in the bound
state and the position of the maximum agree well with the upper stable
fixed point of the deterministic equation. Both depend hardly on the
receptor-ligand distance $\lambda_{wlc}$. The average number of closed
bonds jumps from bound to unbound state in a discontinuous transition.
This transition becomes sharper for larger clusters, that is, the
width of the region in which both states are occupied to an
appreciable degree decreases with increasing cluster size.

The physical reason for the striking plateau in the bound state is the
non-linearity of the worm-like chain force extension relation, which
reflects the strain-stiffening typical for biopolymers. A binding
tether thus pulls the relatively soft transducer until the stiffnesses
match, that is to the regime of harmonic tethers with $\kappa_{wlc} =
1.5$. Therefore, the final extension of the bound tethers decreases less
than linear with $\lambda_{wlc}$ and the effect on the number of
closed bonds is weaker than for the harmonic model. If the transducer
was replaced by another worm-like chain, the behavior of the system
should be more like the harmonic model.

\section{Conclusion}\label{sec:conclusion}

In this paper, we have introduced a stochastic model which allows to
study the interplay of cooperative binding and unbinding for
finite-sized adhesion clusters mediated by tethered ligands. Our model
is based on established principles of receptor-ligand binding,
including Kramers-type rupture rates and separation-dependent binding
rates based on the notion of an encounter complex. By implementing
these principles in the framework of a one-step master equation, we
were able to apply many powerful techniques from stochastic dynamics,
including a Kramers-Moyal derivation of a Fokker-Planck equation
(which in turn corresponds to a stochastic potential) and exact
solutions for stationary solutions and mean first passage times.
In particular, a bifurcation analysis based on the stochastic
potential could be compared to the bifurcation analysis of the
mean field equation to the master equation. 

Our model shows that the simple mechanical model of \fig{cartoon} can
lead to a bistable situation in which two different states of
adhesion, bound and unbound, coexist. The underlying reason for the
occurence of bistability is the existence of two mechanisms for
positive feedback, one for rupture and one for binding. Cooperative
bonds share the force exerted by the transducer so that the force
exerted on each closed bond reduces upon binding. At the same time,
the extension of the tethers reduces upon binding, which then
increases the probability for further binding. In consequence, the
transition rates in the master equation are both strongly non-linear
functions of the number of closed bonds, which both can lead to
positive feedback. This leads to an instability for the intermediate
numbers of closed bonds and thus to bistability. The model discussed
in this paper is an extension to a previously introduced model in
which only the rupture rate was a non-linear function of the number of
closed bonds \cite{uss:erdm04a,uss:erdm04c}. It could be re-obtained
from the current model in the limit of a soft transducer, that is for
$\kappa \to \infty$. In the present discussion, non-linear effects
were mainly due to cooperative rebinding while the non-linearity of
the rupture rate was weak.

In the mean field description, bistability leads to hysteresis for the
binding and unbinding range: binding from the unbound state takes
place at a smaller distance than unbinding from the bound state.
Unlike for previous discussions for non-cooperative bonds
\cite{a:MoreiraEtAl2003,a:MoreiraMarques2004}, this is not due to
kinetic effects. In the stochastic description, the bistable system is
characterized by a bimodal stationary probability distribution.
Fluctuations over the barrier separating the stable states allow
equilibration of the distribution. The time scale for equilibration
has been calculated as the mean first passage time between the
different adhesion states. Due to these finite transition times
kinetic effects from changes of external parameters are important for
the behavior of the system. In analogy to thermodynamic systems at
first order phase transitions, changes of parameters on time scales
smaller than the transition times allow occupation of the metastable
states, while for slower changes, a stationary distribution will
occur. In the bimodal region, the system frequently alternates between
bound and unbound state. For large systems, the transition times
become very large and the transition between bound and unbound becomes
very sharp. The bimodal region in which two states are occupied to an
appreciable and comparable amount becomes very small.

The predictions of this paper for the internal dynamics of adhesion
clusters can be investigated with experimental setups like AFM or the
surface force apparatus which allow to study the specific binding of
two opposing surfaces with controled separation. However, to map out
the occupancy distribution for bound and unbound states could be
tedious because of large transition times and low occupancy outside
the dominant state.  If measuring the binding range by stepwise
reduction of the receptor-ligand distance, binding will take place
when the transition time is smaller or equal to the waiting time. For
large systems this will lead to hysteresis as in the deterministic
case. For smaller systems, hysteresis will decrease due to the smaller
transition times.  In experiments with the surface forces apparatus
\cite{a:WongEtAl1997,a:JeppesenEtAl2001}, a behavior similar to the
one predicted here has been observed. After initial binding a large
attractive force was measured until an equilibrium position of the
surfaces was established. In our model this increase would be due to
the increased extension of the transducer spring. In those
experiments, irreversible bonds with very large affinity have been
used so that repeated transitions between bound and unbound states
could not be observed.

In order to test our results, it would be necessary to use reversible
adhesion bonds with low affinity. In general, binding through
reversible bonds is highly relevant for biological systems.  In
particular, cell-matrix adhesion is mediated by reversible bonds like
the ones between integrin-receptors in the plasma membrane and
fibronectin in the extracellular environment. In this case, the
equilibrium length of the ligand tether is $\ell_b = 11$ nm. Using the
parameter values given in Tab.~I, one then finds from the stochastic
model for $N_t = 5$ that for small adhesion clusters, bistability
should occur around a relaxed receptor-ligand distance of $8$ nm
($\lambda = 0.75$) \cite{uss:erdm06a}.  Together with the ligand and
receptor rest lengths, this results in a cell substrate distance of
around 20 nm, that is the physiological value for cell-matrix
adhesion \cite{a:CohenEtAl2004}.  Therefore the mechanism of
bistability as described here can be used by cells to explore the
extracellular space by many small and transient adhesions. On
encountering favorable conditions, these small adhesions might mature,
e.g.\ by recruitment of additional receptors.  The results presented
here show that this quickly leads to switching times which keep the
adhesions in the bound macrostate.

In order to present a simple and reasonable model, here we have made
the crucial assumption that all bonds are equivalent. This assumption
leads to a one-step master equation and allows the application of many
powerful tools from stochastic dynamics.  In the future, our model
could be extended to include additional aspects of biomimetic or
biological systems, which usually however can not be described in the
framework of a one-step master equation.  In order to describe the
growth and shrinkage of adhesion clusters, the overall number of bonds
$N_t$ should be made dynamic, possible involving regulation through
the cytoskeleton.  Assuming an elastic rather than a rigid transducer
requires solution of elastic equations in order to derive the exact
details of the force distribution. The force distribution would also
be changed when accounting for possible curvature of the opposing
surfaces. Finally it would also be interesting to consider the effect
of disorder, e.g.\ in bond resting lengths or single bond on- and
off-rates.

This work was supported by the German Research Foundation (DFG) through
the Emmy Noether Program and by the Center for Modelling and Simulation
in the Biosciences (BIOMS) at Heidelberg.

%\bibliography{a,books,uss,c,r}
%\bibliographystyle{unsrt}

\end{document}